\documentclass[a4paper,twocolumn]{esapub} 
\usepackage{natbib}
\usepackage{graphicx}

\begin{document}

\title{Irradiance or luminosity changes?}
\author{Sabatino Sofia}
\affil{Department of Astronomy, Yale University, New Haven, CT 06520, USA}
\author[1,2]{Linghuai H. Li}
\affil{Purple Mountain Observatory, Chinese Academy of Science, Nanjing, Jiangsu 210008, China}

\newcommand{\btx}{\textsc{Bib}\TeX}
\newcommand{\filename}{esapub}


\maketitle

\begin{abstract}
Whereas a variation of the solar luminosity, L, will inevitably cause 
a similar change of the total solar irradiance, S, the opposite is not 
true. In fact, the bulk of the days to months variations
of S can be explained entirely in terms of the passage of active regions across the solar
disk. In this case, L remains essentially unchanged.

For the total irradiance variation observed over the solar cycle, 
the issue is more uncertain. One view explains this modulation primarily 
as a combined action of active regions and magnetic network. These
components would be superposed to an otherwise unchanging photosphere.
the other view suggests that the activity cycle modulation of S is 
primarily produced by a variation of L (both in terms of R and 
T$_{\mbox{\scriptsize{eff}}}$) caused by structural reajustments 
of the interior of the Sun induced by a changing magnetic field.
We will present evidence in support of this second interpretation, 
and a model for it. We will also present the S variations over 
the last 5 centuries implied by our model.
\end{abstract}

\section{Introduction}

There is no question that 
\begin{itemize}
\item The hours to months variations of the total irradiance are 
primarily (totally?) due to active regions.
\item The spots depress the irradiance.
\item The faculae add to the irradiance.
\item The contrasts are sufficiently high to be measured with some confidence.
\item The temporal behavior during one rotation is exactly as expected.
\end{itemize}

\begin{figure}
\includegraphics[width=8cm]{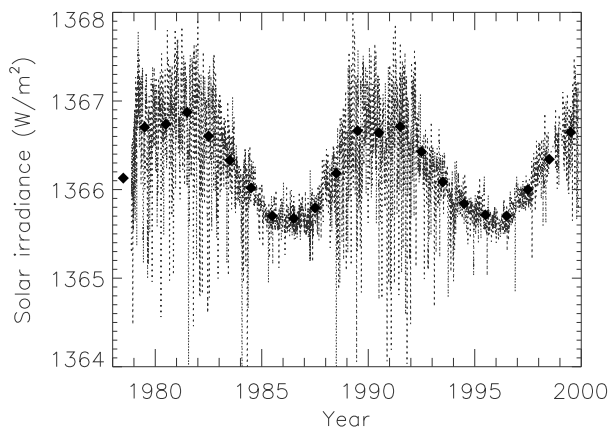}
\caption{
A composite solar irradiance record from the end of 1978 to the present \protect\citep{FL98} and the yearly mean of solar irradiance.
\label{talkfig1}
}
\end{figure}

However, it is not the case for the 11-year cycle, as shown in Figure~\ref{talkfig1} \citep{FL98}.
\begin{itemize}
\item It is usually {\bf assumed} that the magnetic network causes 
  most of the modulation.
\item The measured network contrast is insufficient to account for the 
entire variation \citep{EBF00}.
\item The precision of the irradiance measurements is less certain because 
instrument degradation is more significant than that in short timescales.
\item Proxies of the network are designated, and their magnitude 
is adjusted to minimize residuals with observations.
\end{itemize}
From this viewpoint, it is assumed that the background photosphere 
remains unchanged during the entire cycle.

Of course, an alternative possibility is that most (if not all) 
of the 11 year variability is due to a change in the ''luminosity'' 
without the effects of the magnetic network. In order for that to 
happen, the following are true:
\begin{itemize}
\item The photospheric temperature must change.
\item The internal solar structure must change.
\item The solar radius must change.
\end{itemize}
We propose here the explanation that the 11-year modulation of the 
total irradiance is due to structural adjustments of the solar 
interior in response to a variable internal magnetic field.

\section{Evidence in support of solar structure variations}

\subsection{Variations of solar effective temperature}

\begin{figure}
\includegraphics[width=8cm]{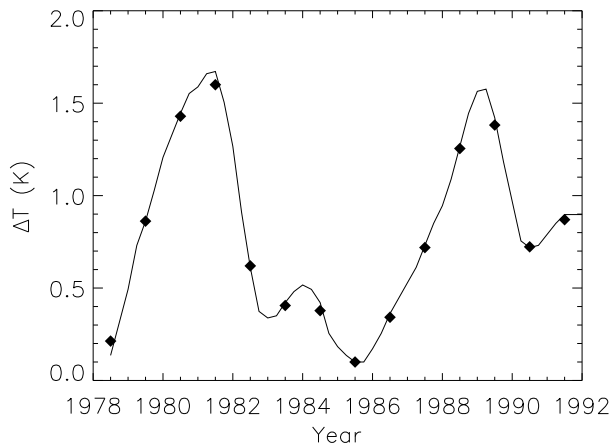}
\caption{
The measured solar photospheric temperature variations from 1978 to 1992 \protect\citep{GL97} and the yearly mean.
\label{talkfig2}
}
\end{figure}

The solar effective temperature was measured by \citet{GL97} 
from ratios of spectral line depths of
\[
  \mbox{C\,I($\lambda 5380$)/Fe\,I($\lambda 5379$)}
\]
and
\[
  \mbox{C\,I($\lambda 5380$)/Ti\,II($\lambda 5381$)}
\]
The excitation potentials of these lines are different from each other.
\[
  \mbox{C\,I$ = 7.68$ eV},
\]
\[
  \mbox{Fe\,I$ = 3.69$ eV},
\]
\[
  \mbox{Ti\,II$ = 1.57$ eV}.
\]
The consistency of results indicates that the $T_{\mbox{\scriptsize{eff}}}$
they measure is photospheric temperature.
The spectroscopic temperature variations of the sun measured by Gray and Livingston 
(1997) over the period from 1978 to 1992, are shown in Figure~\ref{talkfig2}.
The zero point is chosen arbitrarily.

\subsection{Variations of solar oscillations}

Solar-cycle effects on solar oscillation frequencies 
were determined by \citet{LW90}. Recently, \citet{BJT99} 
presented a correlation analysis of GONG p-mode frequencies 
with nine solar activity indices for the period from 1995 
August to 1997 August. A decrease of 0.06 $\mu$Hz in frequency during 
the descending phase of solar cycle 22 and an increase of 0.04 $\mu$Hz 
in the ascending phase of solar cycle 23 are observed. These results 
provide the first evidence for change in p-mode frequencies around 
the declining phase of cycle 22 and the beginning of new cycle 23.
This analysis further confirms that the temporal behavior of the 
solar frequency shifts closely follow the phase of the solar activity cycle.
Besides, the analysis given by \citet{HKH99} suggests that the solar 
cycle related variation of the oscillation frequencies is not due to 
contamination of observed Doppler shifts by the surface magnetic fields. 

\subsection{Radius variations}

Ground-based measurements of the solar radius exist over three 
centuries, but the results are controversial and inconsistent.
When a homogenized data base covering observations over the 
last three centuries is used, \citet{B98} found a statistically 
significant positive correlation between solar radius and 
sunspot numbers. Measurements of the solar radius made with 
the Danjon astrolabe at Santiago, Chile, and with the 
magnetograph of the solar telescope of Mount Wilson 
Observatory during the period 1990-1995, show similar 
variations in time and with a similar trend as the 
variation of sunspot numbers \citep{N97}.

The space-based MDI-SOHO limb observations \citep{EKBS00}
also show that the cycle variation of the solar radius is 
in phase with sunspot numbers. However, the estimated upper limit for 
the cycle variation is 
$\delta r_{\mbox{\scriptsize{cycle}}} = 21\pm 3$ milliarcsec.

All the above are inconsistent with an unchanging solar 
interior and suggest changes within the solar interior.

\section{Method}

Several years ago, \citet{EST85} proposed that a variable 
internal magnetic field should affect all the global 
parameters of the sun. Subsequently, \citet{LS95} 
carefully systematized the formulation of the problem, 
and wrote a code to do exploratory calculations. They 
found that sensible internal magnetic fields variations 
would perturb he internal structure of the sun, and 
consequently affect all global solar parameters.

The formulation and the code was further generalized 
by \citet{LS00}, and it is still being enriched at the present 
time. Elements of the new code are:
\begin{itemize}
 \item Include the magnetic energy per unit mass $\chi$, 
and the ratio of magnetic pressure to magnetic energy, $\gamma-1$,
as two additional variables in stellar structure and evolution.
\item Take into account influence of magnetic fields on radiative opacities.
\item Take into account all time-dependent contributions to the 
equations of stellar structure (we need short timescales).
\item Modify the radiative loss assumption of a convective element to 
include local turbulence effects associated with small-scale magnetic 
fields.
\item Use real equations of state on computing first and second order 
derivatives associated with magnetic fields.
\item Use the most up-to-date stellar evolution codes (YREC7) since 
the effects we wish to determine are very small.
\end{itemize}

\section{Results}

\begin{figure}
\includegraphics[width=8cm]{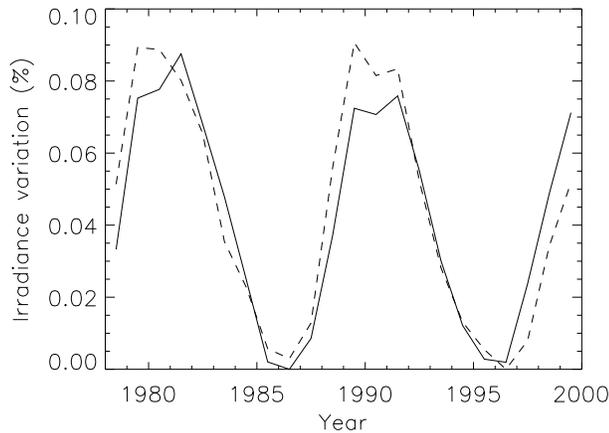}
\caption{
Comparison between the measured (solid curve) and calculated (dashed curves) solar irradiance variations.
\label{talkfig3}
}
\end{figure}

\begin{figure}
\includegraphics[width=8cm]{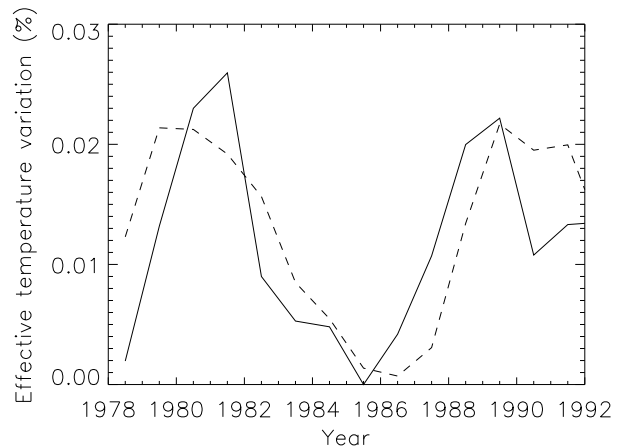}
\caption{
Comparison between the measured (solid curve) and calculated (dashed curve) solar photospheric temperature variations.
\label{talkfig4}
}
\end{figure}

\begin{figure}
\includegraphics[width=8cm]{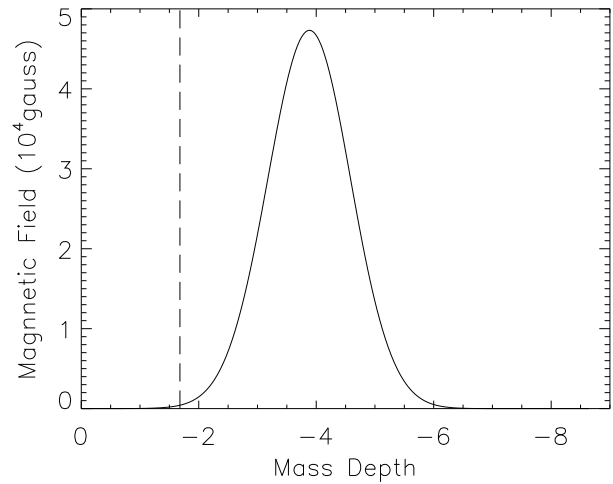}
\caption{
Distribution of the inferred magnetic field in the solar interior in 1989 according to the measured irradiance and photospheric temperature cyclic variations given in Figures~\ref{talkfig1} and~\ref{talkfig2}. The vertical line indicates the base of the convection zone.
\label{talkfig5}
}
\end{figure}

\begin{figure}
\includegraphics[width=8cm]{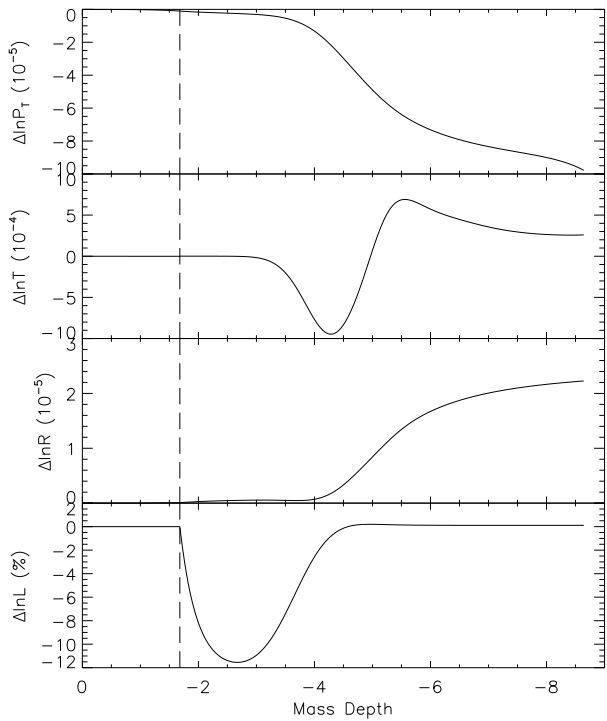}
\caption{
The structural changes caused by the magnetic field distributions given in Figure~\protect\ref{talkfig5}: relative pressure, temperature, radius and luminosity changes from top to bottom. The vertical line indicates the base of the convection zone.
\label{talkfig6}
}
\end{figure}

We use this code to show that the entire 11-year variations of the total irradiance (\ref{talkfig3}) 
and $T_{\mbox{\scriptsize{eff}}}$ (see Figure~\ref{talkfig4}) could 
be produced by a magnetic field of strength (20-47 kG) and location ($r=0.96 R_{\odot}$)
equal to that determined from helioseismology \citep{ACT00}, as shown in 
Figure~\ref{talkfig5}. Figure~\ref{talkfig6} shows the corresponding internal structure adjustment of the sun.
The calculated cycle change of the solar radius is about $0^{\prime\prime}.02$, which is in agreement with the MDI/SOHO observation \citep{EKBS00}.

From this fit we find that the maximum magnetic field 
in the solar interior, $B_m$, is related to $R_Z$ via
\begin{equation}
  B_m = B_0 \{190 + [1 + \log_{10}(1+R_Z)]^5\}, \label{bm}
\end{equation}
where $R_Z$ is the yearly-averaged sunspot number, $B_0=90$ G. The profile of the magnetic energy per unit mass $\chi$ is descibed by a gaussian function
\begin{equation}
  \chi=\chi_m\exp[-\frac{1}{2}(M_D-M_{\mbox{\scriptsize{Dc}}})^2/\sigma^2], \label{gauss}
\end{equation}
where $M_{\mbox{\scriptsize{Dc}}}=-4.25$ specifies the location and $\sigma=0.5$
specifies its width. $B_m$ is used to determine $\chi_m$. The mass depth $M_D$ is defined as
\[
  M_D = \log_{10}(1-M_r/M_{\odot}).
\]

\begin{figure}
  \includegraphics[width=8cm]{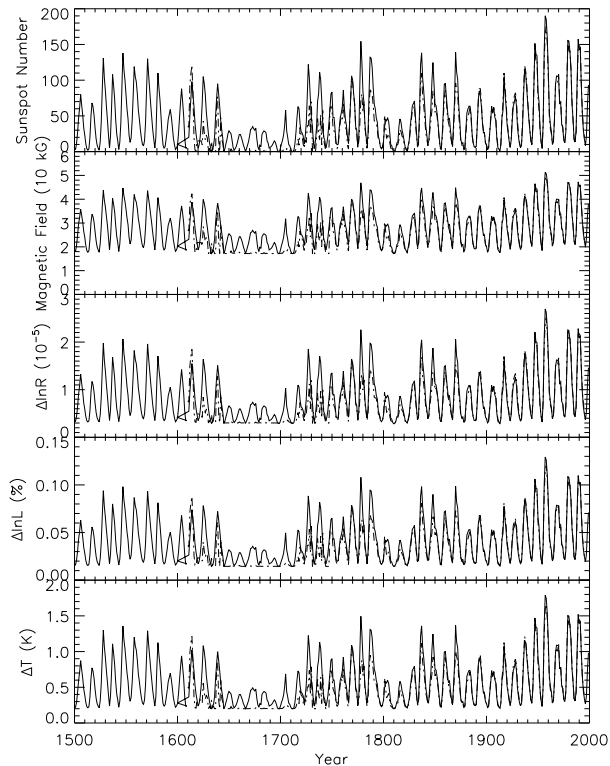}
\caption{Solar variability in the past five centuries (solid curve corresponds to the Z\"{u}rich sunspot number $R_Z$) and in the past four centuries (dashed curve conrresponds to the group sunspot number $R_G$).
kG stands for kilo-Gauss, $L$ for total solar luminosity, $T$ for solar effective temperature, $R$ for solar radius.
}
\label{talkfig7}
\end{figure}

Using Eqs.~(\ref{bm}) and (\ref{gauss}), we can extrapolate the solar irradiance back during the period when the annual sunspot numbers are available \citep{S83,HS98}, as shown in Figure~\ref{talkfig7}.

As we can see from Fig.~\ref{talkfig7}, the maximum variability of the solar radius is about $2\times 10^{-5}$, or 0.02 arc~s.  Although this variation is in agreement with the most recent determination of the cycle radius variations obtained from the MDI experiment on SOHO \citep{EKBS00}, it is much smaller than the radius changes determined from historical data over the last 2 centuries. In our view, the most reliable historical data sets from which solar radius changes can be determined are the duration of total eclipses measured near the edges of totality. From them, changes of the order of 0.5 arc~s have been detected.  In particular, a change of $0.34$ arc~s between 1715 and 1979 \citep{DSFHM80}, a change of $0.5$ arc s between 1925 and 1979 \citep{DSFHM80}, and no change between 1979 and 1976 \citep{SDDF83}, were detected. If such changes are real, what could cause them? What are the corresponding solar irradiance changes?

If we use the magnetic field location required to produce the 11 year cycle variability, we find that it is impossible to produce a 0.5 arcsec radius variation even if we apply an unreasonablely strong magnetic field. However, our model shows that the deeper the location of a magnetic field and the more intense the magnetic field, the larger the resulting radius change.

\begin{figure}
  \includegraphics[width=8cm]{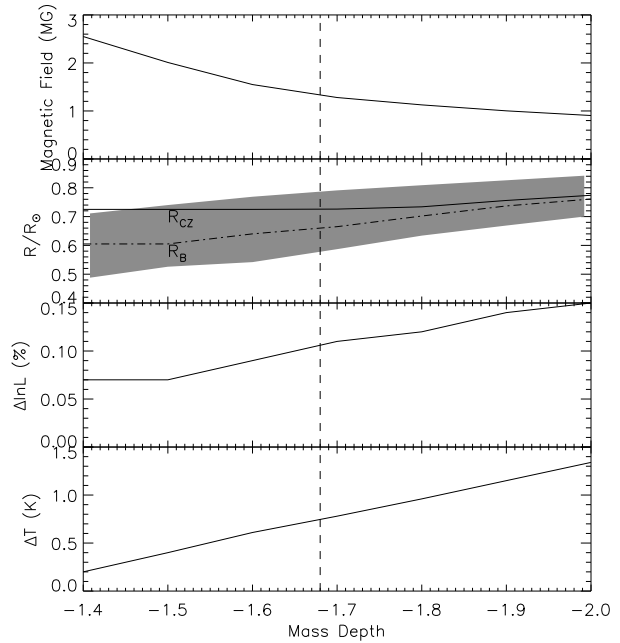}
\caption{Solar variability corresponding to the probable change in the solar radius between 1715 to 1979.
$R_{\mbox{\scriptsize{CZ}}}$ is the location of the base of the convection zone, while $R_B$ is the location of maximal magnetic field. MG stands for Million Gauss. The mass depth is defined as $\log(1-M/M_{\odot})$ by the mass coordinate M. The smaller the mass depth, the closer to the surface of the model sun.
}
\label{talkfig8}
\end{figure}

We thus compute the magnetic field required to produce the detected radius change between 1715 and 1979 as a function of mass depth, as displayed in the top panel in Fig.~\ref{talkfig8}. It is well known that a strong magnetic field will cause a change of location of the boundary between the convective and the radiative region \citep{LS95}. The second panel from the top in this figure shows how the convection boundary $R_{\mbox{\scriptsize{CZ}}}$ varies with the applied magnetic field (solid curve), and how the location of the maximal magnetic field, $R_B$, varies with the mass depth (dashed curve). The shadowed region indicates the half-width of the required magnetic field. Of particular relevance are the values corresponding to the base of the convection zone, as indicated by the dot-dashed line in this figure, since all conventional dynamo models locate the process precisely at that depth. There, the magnetic field required to cause a 0.34 arcsec change of the solar radius is 1.3 million G, and the resulting luminosity variation is 0.12 percent (the third panel of Fig.~\ref{talkfig8}), which is almost equally due to the variation of effective temperature (the bottom panel) and radius, since the radius variation contributes $2\times \Delta\ln R=0.07\%$. These values are interesting for producing significant climate change if the solar variations are sufficiently long lasting, and for not grossly contradicting what we know about the Sun, excepting a value for the magnetic field that is larger than we are comfortable with, but it is not in conflict with helioseismology \citep{ACT00,SL00}.

\section{Conclusions}

From what we present above, we reach the following conclusions:
\begin{itemize}
  \item The total irradiance variation, and the photospheric 
temperature variation observed over the 11-year activity 
cycle can be explained in terms of the variation of an 
internal solar magnetic field of 20-47 kG located at 
$r=0.96 R_{\odot}$.
  \item The above result is in agreement with helioseismological 
data, and with the variations of the solar radius measured with 
MDI/SOHO.
  \item The extrapolation of this process to the past 2-3 
centuries produces a change in luminosity of only 0.1\%, 
and a radius change of only $0.02$ arcsec.
  \item If radius variations of order $0.5$ arcsec do occur, 
a larger (1.3 MG) variation of a field located below 
the base of the convection zone is required.
  \item The combined effect of both phenomena can yield 
a $\Delta L$ of 0.2\%\ over many decades.
\end{itemize}

\section*{Acknowledgments}

This work was supported in part by a NASA grant and Li was supported by
Natural Science Foundation of China (project 19675064).

\begin{thebibliography}{}

\bibitem[Fr\"ohlich \&\ Lean(1998)]{FL98}
 Fr\"ohlich, C. \&\ Lean, J. 1998, in IAU Sympsium 185: 
 New Eyes to See Inside the Sun and Stars, ed. F.L. Deubner, 
 (Dortrecht: Kluwer Academic Publ.), 89

\bibitem[Ermolli et al.(2000)]{EBF00}
Ermolli, I., Berrilli, F., Florio, A. 2000, this proceedings

\bibitem[Gray \&\ Livingston(1997)]{GL97}
Gray, D.F. \&\ Livingston, W.C. 1997, ApJ, 474, 802

\bibitem[Libbrecht and Woodard(1990)]{LW90}
Libbrecht, K. G., Woodard, M. F. 1990, Nature, 345, 779

\bibitem[Bhatnagar et al.(1999)]{BJT99}
Bhatnagar, A., Jain, K.,  Tripathy, S. C., 1999, ApJ 521, 885

\bibitem[Howe et al.(1999)]{HKH99}Howe, 
R., Komm, R.,Hill, F., 1999, ApJ 524, 1084

\bibitem[Basu(1998)]{B98}
Basu, D. 1998, Solar Phys., 183, 291

\bibitem[No\"{e}l(1997)]{N97}
No\"{e}l, F. 1997, A\&A, 325, 825

\bibitem[Emilio et al.(2000)]{EKBS00}
Emilio, M., Kuhn, J.R., Bush, R.I. \&\ Scherrer, P. 2000, ApJL, in press

\bibitem[Endal et al.(1985)]{EST85}
Endal, A.S., Sofia, S. \&\ Twigg, L.W. 1985, ApJ, 290, 748

\bibitem[Lydon \&\ Sofia(1995)]{LS95}
 Lydon, T.J. \&\ Sofia, S. 1995, ApJ Suppl., 101, 357

\bibitem[Li and Sofia(2000)]{LS00}
Li, L. H., Sofia, S., submitted to ApJ (see astro-ph/0007203).

\bibitem[Antia et al.(2000)]{ACT00}
 Antia, H.M., Chitre, S.M., \&\ Thompson, M.J. 2000, A\&A, in press

\bibitem[Schove(1983)]{S83}
Schove, D. J., Sunspot cycles (Stroudsburg: Hutchinson Ross Publ., 1983), 10
\bibitem[Hoyt \&\ Schatten (1998)]{HS98}
Hoyt, D. V., Schatten, K. H. 1998 Solar Phys. 181, 491

\bibitem[Dunham et al.(1980)]{DSFHM80}Dunham, D. W., Safia, S., Fiala, A. D., Herald, D., Muller, P. M. 1980, Science, 210, 1243
\bibitem[Sofia et al.(1983)]{SDDF83}Sofia, S., Dunham, D. W., Dunham, J. B., Fiala, A. D. 1983,  Nature, 304, 522

\bibitem[Sofia \&\ Li(2000)]{SL00}
Sofia, S. \&\ Li, L. H., 2000, submitted to JGR


\end{thebibliography}
\end{document}